\documentstyle[prl,aps,epsf]{revtex}
\draft
\begin{document}
\twocolumn[\hsize\textwidth\columnwidth\hsize\csname @twocolumnfalse\endcsname

\title{Temperature and ac Effects on Charge Transport in Metallic Arrays
of Dots} 
\author{C. Reichhardt and C.J. Olson Reichhardt} 
\address{ 
Center for Nonlinear Studies and Theoretical Division, 
Los Alamos National Laboratory, Los Alamos, New Mexico 87545}

\date{\today}
\maketitle
\begin{abstract}
We investigate the effects of 
finite temperature, dc pulse, and ac drives on the charge transport in 
metallic arrays using numerical simulations. For  
finite temperatures there is a finite conduction threshold 
which decreases linearly with temperature.
Additionally we find 
a quadratic scaling of the current-voltage curves
which is independent of temperature for finite thresholds. 
These results are in 
excellent agreement with recent experiments on 2D metallic dot arrays. 
We have also investigated the effects of an ac drive as well as
a suddenly applied dc drive. 
With an  ac drive the conduction threshold decreases for fixed frequency and
increasing amplitude and saturates for fixed 
amplitude and increasing frequency. For sudden applied dc drives below 
threshold we observe a long time power law conduction decay. 
\end{abstract}

\vskip2pc]
\narrowtext
The continuing push toward miniaturization of electronics has
led to considerable 
interest in the conductivity properties of arrays of nanoscale
metallic dots.
Previous studies 
of charge transport in arrays of metallic dots have established that there
can be a finite voltage threshold for conduction 
\cite{Wingreen,Rimberg}.
Additionally the current-voltage curves can be nonlinear 
and exhibit scaling such as that observed for dynamic
critical phenomena \cite{Fisher}.
Middleton and Wingreen (MW) studied charge transport in 1D and 2D arrays
with both simulations and analytic theory 
\cite{Wingreen}. They find scaling in the I-V curves for 1D arrays 
of the form $I = (V/V_{T} - 1)^\zeta$
with
$\zeta = 1.0$, while for the 2D system       
they predict $\zeta = 5/3$ and find in simulations $\zeta = 2.0$.
Recent simulations by Reichhardt and Olson (RO) 
\cite{Reichhardt} also produced similar scaling
exponents as well as a crossover from 2D disordered filamentary charge flow
regions to more ordered 1D flowing channels for increasing drive.
Experimental studies of metal dot arrays have also found 
scaling in the IV curves for 2D and 1D systems 
\cite{Rimberg,Duroz,Bezyryadin,Black,Kurdack,Jaeger,Clarke,Wybourne,Lebreton,Ancona};
however,   
the scaling exponents in these experiments exhibit a wide range of
values from $\zeta=1.4$ to $2.5$.
The spread in the exponents may be due to 
different types of disorder present in these arrays. 

Recent experiments by 
Parthasarathy {\it et al.} \cite{Jaeger} 
have specifically sought to address the role of different types
of array 
disorder on the current-voltage scaling. 
They considered varying 
the global structural order of the array by varying the amount of voids.  
For structurally ordered systems composed of
a triangular monolayer 
of gold nanocrystals without voids, they observe a single power law scaling
with $\zeta=2.25$ in the current-voltage curves.
In this case there is
still charge disorder in the substrate and
disorder in the inter-particle couplings.
For arrays where structural disorder is added,
the current-voltage curves 
could not be fit by a single power law. 
Simulations by RO for structurally disordered 
arrays also produced 
similar behaviors. These results suggest that differences
in structural disorder in the earlier experiments may be the cause 
of the differences in the observed exponents.

Less is known about how other types of disorder, such as thermal
disorder, or
perturbations, such as ac drives, would affect the scaling or the
exponents.
Recently Parthasarathy {\it et al.} \cite{Jaeger2} have investigated the
role of finite temperature 
on the current-voltage curves for ordered and structurally disordered 
gold nanoparticle arrays. They find for the ordered arrays that the
threshold voltage decreases linearly with $T$ while the scaling exponent
is unaffected. For the disordered arrays the threshold also decreases linearly
at low $T$.  Decreases in the threshold with increasing
temperature have also been observed in previous experiments on disordered
arrays 
\cite{Lebreton,Ancona,ClarkeAPL,Pepin,Cordan00}.
In both the ordered and disordered arrays, for 
higher temperatures the threshold is lost and the nonlinear scaling of the
IV curve disappears and is replaced by linear behavior.
Other experiments on ordered arrays have also found that temperature 
only weakly affects the shape of the IV curves 
\cite{Bezyryadin,Jaeger}. 
Previous simulations have been limited to $T = 0$ \cite{Wingreen,Reichhardt} 
or have considered only very small 
arrays \cite{Cordan00,Leroy}.

Another type of perturbation that has recently been considered 
in experiments on highly resistive samples
is the application of a sudden dc drive \cite{Morgan}. 
The application of a dc voltage that is below the conduction threshold
voltage produces a current response that shows a two stage decay. 
The first stage, shown in Fig.~4 of Ref. \cite{Morgan}, 
is a rapid decay of the current at short time scales
that does not fit to a power law. 
For longer times, however, the current shows a power law decay with
$I(t) \propto t^{-\alpha}$, with $0.1 < \alpha < 1.0$, depending 
on the applied voltage and temperature.  

A third perturbation which can also be applied to the system is
an ac drive.
To our knowledge, the effect of ac drives on the IV curves 
has not been investigated by simulation, nor has it been considered
in experiments.
It is not clear whether increasing ac amplitudes may cause the 
scaling of the IV curves to be lost, as the perturbation 
%

\begin{figure}
\center{
\epsfxsize=3.5in
\epsfbox{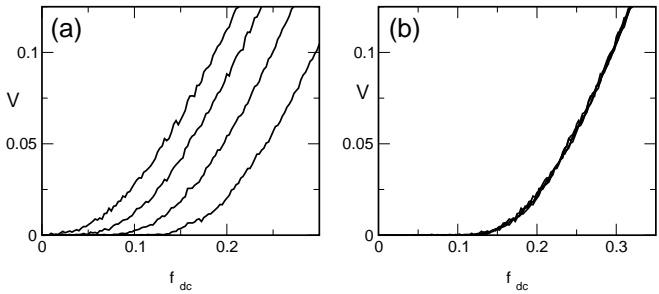}}
\caption{
(a) The velocity $V$ vs dc driving force $f_{dc}$ for (from
right to left) $T/T_{th} = 0.0, 0.24, 0.61$ and $0.95$, 
where $T_{th}$ is the 
temperature at which the threshold disappears, $f_{th}=0$. 
(b) The same curves collapsed
by applying a linear shift of the $x$-axis, $f_{\rm shift}(T)$.  
}
\label{fig:scale}
\end{figure}

\noindent
due to
temperature did.  

In this paper we use the RO model for charge transport through metal dot
arrays to consider the effect of finite temperature
and ac drives on the threshold behavior and the current-voltage scaling. 
We also consider the case of sudden applied dc drives and examine the
conduction decay.
For this work we consider only 2D square arrays, and use system
sizes from $20\times 20$ to $60\times 60$. Our results are mainly presented
for system sizes of $50\times 50$ which we previously found to 
be adequately large to capture the essential physics.  
The sample has periodic boundary conditions in 
the $x$ and $y$-directions and contains $N_{c}$ mobile charges. 
The equation of motion for a charge $i$ is  
\begin{equation}
{\bf f}_{i} = \eta{\bf v}_{i} = -\sum_{j}^{N_{c}}\nabla U(r_{ij})
 + {\bf f}_{p} + 
{\bf f}_{dc}  + {\bf f}^{T} + {\bf f}_{ac}.
\end{equation}

 The mobile charges
interact with a Coulomb term, $U(r) = q/r$. We employ
a fast summation technique for
computational efficiency to calculate the long-range
Coulomb force \cite{Jensen}. 
The dc driving term is ${\bf f}_{dc} = f_{dc}{\hat {\bf x}}$ which would 
arise from a dc applied 
voltage $V$. On each plaquette there is a threshold force 
$f_{p}$, chosen from a Gaussian distribution, which prevents the charge from
leaving the plaquette until $f_p$ is exceeded.  This threshold
originates from the energy required
to add an electron  to a dot with charge $q$.
The threshold for dot $j$ 
is $V_{th}^{j} = q_{j}/C_{j}$, where $C_j$ is the capacitance of 
dot $j$. We measure the global charge flow or current 
$I = V_{x} = (1/N_{c})\sum{\bf v}_{i}\cdot{\bf {\hat x}}$.
Starting from zero we increase the dc drive in increments. 
When measuring I-V curves, we wait
at each increment for 1500 simulation steps 
before taking data to avoid transient
effects. We study transient effects separately in Section II of
this paper.
To explore finite temperature effects, we add a thermal 
force term ${\bf f}^{T}$ which has the properties 
$<f^{T}(t)> = 0$ and $<f^{T}_i(t)f^{T}_j(t^{\prime})> = 
2\eta k_{B} T \delta_{ij}\delta(t-t^{\prime})$.
Here $\eta$ is a damping constant which we set equal to unity. 
The damping  
corresponds to dissipation produced by the motion of the charge. 
In Section III of this paper, we also add a term
representing an external ac drive, ${\bf f}_{ac}
= A\sin(\omega t)$, which 

\begin{figure}
\center{
\epsfxsize=3.5in
\epsfbox{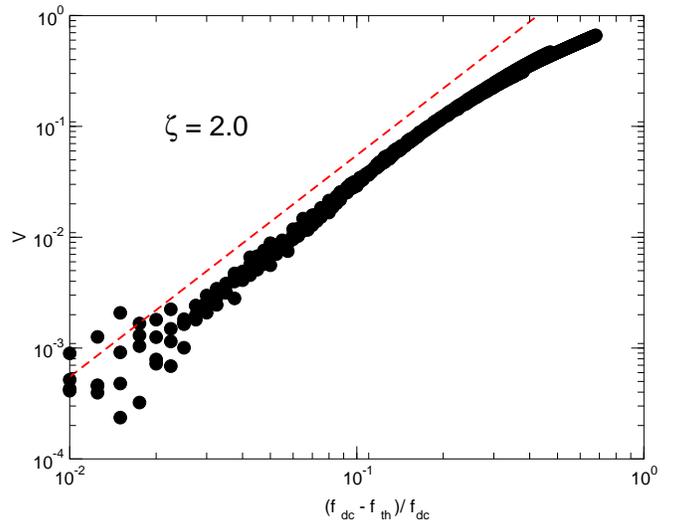}}
\caption{
The scaling of the average velocity $V$ vs applied drive for 
the curves in Fig.~1(b). The dashed line indicates $\zeta=2$.
}
\label{fig:nu2}
\end{figure}

\noindent
would arise from an applied
ac voltage. Here $A$ is the amplitude and $\omega$ is the frequency
of the ac drive.        

\section{Temperature effects}
\begin{figure}
\center{
\epsfxsize=3.5in
\epsfbox{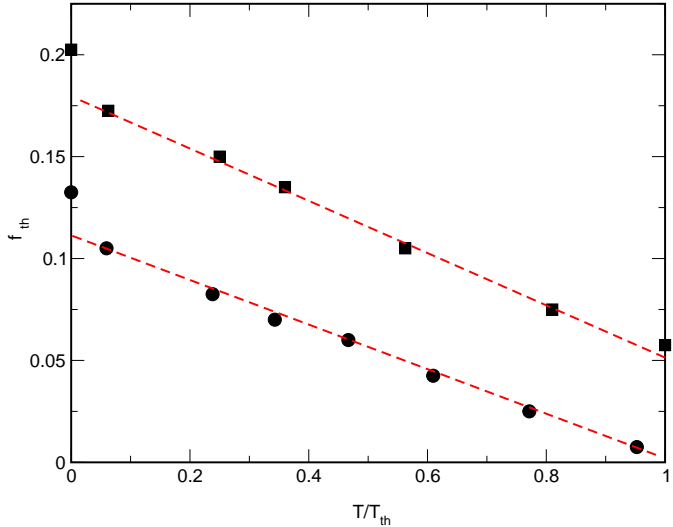}}
\caption{
The threshold $f_{th}$ vs T for (circles) the system in Fig.~1 with
$f_{p} = 4.0$ and (squares) a system with $f_{p} = 8.0$. The solid curves
are linear fits. For both sets the temperature axis is normalized by $T_{th},$
the temperature at which there is no threshold for the $f_{p} = 4.0$ system. 
}
\label{fig:ft}
\end{figure}

We first consider the case of different temperatures and zero ac drive. 
We normalize our temperature in units of $T_{th}$ which is the temperature at
which the threshold force for motion, $f_{th}$, becomes 
zero.  In Fig.~\ref{fig:scale}(a) we plot the 
velocity-force curves (current-voltage curves) 
for a sample with $f_{p}=4.0$ 
for increasing temperature,
$T/T_{th}=$ 0, 0.24, 0.61, and 0.95,
indicating that the finite temperature driving threshold $f_{th}$ 
decreases with
temperature. In Fig.~\ref{fig:scale}(b) 
we show that the curves can be collapsed in 
the same manner as the experimental curves in Ref. \cite{Jaeger2},
by linearly shifting the $x$-axis an amount given by
$f_{\rm shift}(T)=f_{th}(T)-f_{th}(0)$.
This collapse shows that the 
scaling exponent is {\it independent} of temperature.  
In Fig.~\ref{fig:nu2}, a log-log plot
of $V$ vs $(f_{dc} - f_{th})/f_{th}$ for the 
curves in Fig.~\ref{fig:scale}(b)  
illustrates a power law scaling with $\zeta = 2.0 \pm 0.15$, 
in good agreement
with the experimental values \cite{Jaeger,Jaeger2}.   
In addition, the thresholds are decreasing linearly with temperature,
as indicated in Fig.~\ref{fig:ft}.
If the temperature is further increased above 
$T/T_{th}=1$, the threshold
velocity disappears and the shape of the I-V curve begins to change.
Thus above this temperature
it is no longer possible to rescale the I-V curves by a simple shift of
the $x$-axis.  This is also in agreement with
experiment \cite{Jaeger2}.

In Fig.~\ref{fig:ft} we show the conduction threshold 
$f_{t}$ vs $T$ for two systems that
have different average disorder strength,
$f_{p}=4.0$ and $f_{p}=8.0$. Both sets are normalized by $T_{th}=1.0$, the
temperature at which the threshold reaches 
zero for the $f_{p} = 4.0$ system (circles). 
Here the 

\noindent
thresholds decrease
linearly with temperature for all but the lowest temperatures,
$T/T_{th} < 0.1$. For the sample with $f_{p} = 8.0$ (squares), the overall 
thresholds are higher which is consistent with the increased average
force to leave a plaquette. 
The linear decrease in the threshold is   
in agreement with experimental observations \cite{Jaeger2}.
We have tested 
several different methods for determining the threshold, such 
as using different finite velocity percentages ranging from
$0.005$ to $0.10$, 
and find consistent linear decreases in the
threshold with temperature.

The fact that the scaling of the I-V curves changes with increasing
temperature only once a threshold temperature $T_{th}$ has been
exceeded can be understood with a simple physical picture of the
channels of charge motion.  It was shown previously at $T=0$ that, for low
applied voltages, the charges move through riverlike patterns of
channels inside the sample \cite{Reichhardt}.  The exact pattern of
the channels in a given sample is determined by the specific
realization of disorder within that sample.  Throughout the nonlinear segment
of the I-V curve, charge motion is confined to a number of channels
that increases as $f_{dc}$ is increased.  Channels with the lowest
barrier to motion open first, followed by channels with increasing
barriers to motion.  The order in which the channels open is fixed by
the disorder realization.  When temperature is applied, the barrier
to motion in each channel is effectively lowered; however, the relative
barrier heights of the channels are unchanged.  Thus, at low but
finite temperatures, the channels open in the same order as at $T=0$,
but at reduced values of $f_{dc}$.  Therefore the shape of the I-V
curve is merely shifted to lower values of $f_{th}$, but not altered.
In contrast, once the temperature is increased enough to completely
eliminate the barrier to motion in some of 

\begin{figure}
\center{
\epsfxsize=3.5in
\epsfbox{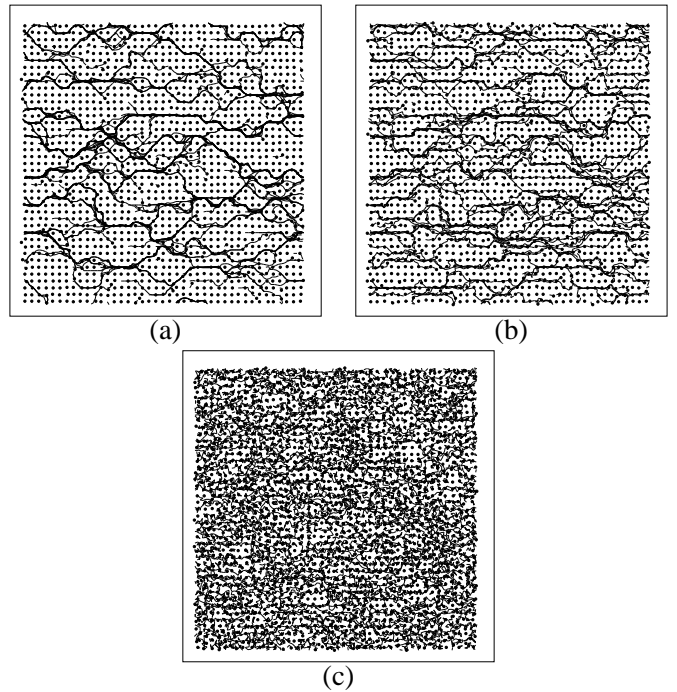}}
\caption{Images of the charge flow channels at fixed
$f_{dc}=0.125$ and different temperatures.
(a) $T/T_{th}=0$; (b) $T/T_{th}=0.25$; (c) $T/T_{th}=1.0$.
The total flux of charge through the sample is the same for
each image.
}
\label{fig:rivers}
\end{figure}

\noindent
the channels, all of these
channels open immediately and the order in which the channels open
is changed, which amounts to a change in the topography of the
channel structure \cite{River}.  
This alters the shape of the I-V curve and causes 
the scaling to be lost, in addition to producing $f_{th}=0$.

To illustrate this picture, in Fig.~\ref{fig:rivers} we show the
flow pattern of the charges in our simulation at three different 
temperatures for fixed $f_{dc}=0.125$.
In Fig.~\ref{fig:rivers}(a) and (b), the temperatures $T/T_{th}=0$ and
$T/T_{th}=0.25$ are well below the crossover temperature, 
and linear scaling of the I-V curves is still possible.  
Although the details of the smallest channels vary slightly,
the overall pattern of the primary channels is the same in both panels.
In contrast, Fig.~\ref{fig:rivers}(c) 
shows the channel structure at threshold, $T/T_{th}=1$.  
Here, although there is still inhomogeneous flow, 
the channel pattern seen at lower temperatures has been destroyed.

If this picture correctly captures the behavior,
it implies several experimentally testable features.  First, the 
strength of the disorder will determine the persistence of the channel
patterns.  In samples with stronger disorder, the threshold temperature
should increase, as in Fig.~\ref{fig:ft}.  
For metallic dots, the disorder strength is determined by the
inverse capacitance of each dot, which goes
as $C=4\pi\epsilon\epsilon_0 r$.  Thus, samples containing dots with
smaller radii should show a higher threshold temperature.
Secondly, the topography of the channels of charge flow is strongly
correlated with the noise in the charge velocity, 
\begin{figure}
\center{
\epsfxsize=3.5in
\epsfbox{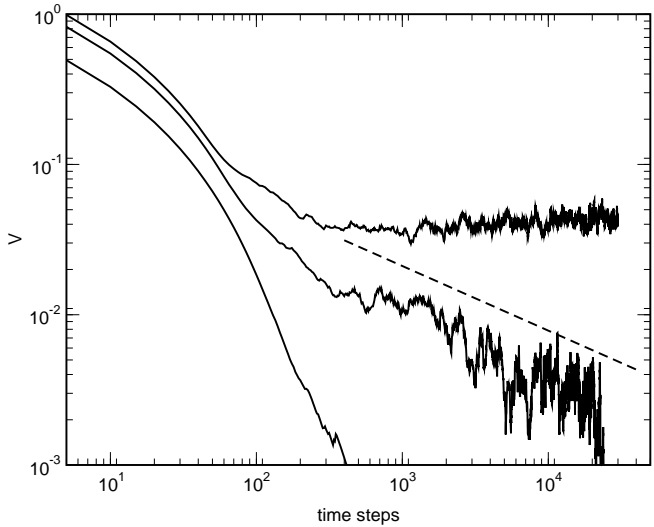}}
\caption{  
The velocity vs time for a sudden applied dc drive 
for the system in Fig.~1 at $T = 0.0$.  
From top to bottom, 
$f_{dc}/f_{th} = 0.95$, $0.65$, and $0.35$. The dashed line is 
a power law with $\alpha = 0.45$. 
}
\label{fig:trans}
\end{figure}

\noindent
as
has been demonstrated for 
the case of superconducting vortices \cite{River,Volt}.
If the structure of the channels changes 
above the threshold temperature, this should be observable as
a change in the noise characteristics.  
Finally, in samples that contain
voids, the channels are highly constrained by the voids themselves, and
the channel pattern cannot be altered even by temperature.  
Thus, in heavily voided arrays,
the linear scaling behavior of the I-V curves should hold to much higher
temperatures than in similar void-free arrays.

\section{Transient effects}

We next consider the case where the dc drive is not slowly increased 
from zero, but is instead instantly increased to a constant value. 
Here, transient charge flows are possible
when the applied force is stronger than the local threshold for 
certain regions of the sample, 
even though the force is below the global threshold of the array.
We note that when the drive is slowly increased we do 
sometimes observe very small 
charge rearrangements at the onset of a force increment.
As in  recent experiments \cite{Morgan,Ginger},
we measure the transient velocity response $V(t)$ as a function of time.
We
focus on the $T = 0$ case for the system with $f_p=4.0$ that was 
shown in Fig.~\ref{fig:scale}. 
In Fig.~\ref{fig:trans} we plot $V$ vs time curves for sudden applied
drives of $f_{dc}/f_{th} = 0.95, 0.65,$ and $0.35$
that are indicative of the three types of decays we observe.
At low drives $f_{dc}/f_{th}<0.5$, 
as illustrated by $f_{dc}/f_{th} = 0.35$ (bottom),
the velocity response drops quickly to zero. 
In this case, although some charge rearrangement occurs throughout
the dots, none of the charges moves more than a lattice constant
and there
are no correlated riverlike structures 

\begin{figure}
\center{
\epsfxsize=3.5in
\epsfbox{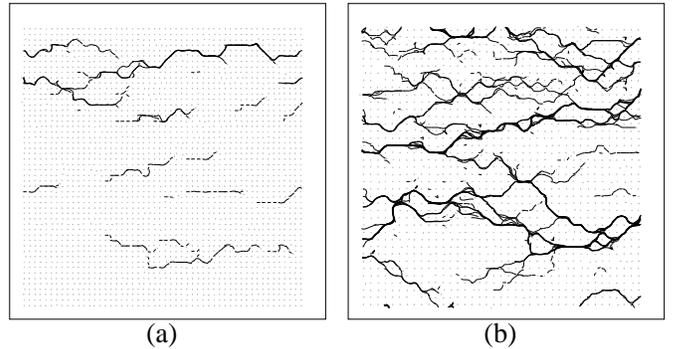}}
\caption{
(a) Flow channels for 
the middle curve in Fig.~5 at $f_{dc}/f_{th}=0.65$
showing channels moving only for 
a finite time, then stopping. (b) Flow channels for the top curve in Fig.~5
at $f_{dc}/f_{th}=0.95$.
Here the channels wrap around the periodic boundary conditions
and the flow does not come to a stop.  
}
\label{fig:trans2}
\end{figure}

\noindent
of motion.
For higher
drives such as $f_{dc}/f_{th} = 0.65$ (middle) we observe  
two regimes: a short time decay that does not have a power law form,
and a slow decay at longer times that fits well to a power law 
$V \propto t^{\alpha}$ with 
$ \alpha = -0.45$ (indicated by the dashed line). 
For the short time decay 
we again observe individual charge rearrangements throughout the
array which quickly settle down. For 
the longer times we observe correlated river structures
such as those illustrated in Fig.~4(a).  The
river structures are smaller than the system
size and the rivers die out as time passes.
In Fig.~6(a) we show the flow patterns for 
$f_{dc}/f_{th}=0.65$ (middle curve of Fig.~5)
where some 1D winding channels
of finite length occur in different places
in the sample. 

For higher drives,  as illustrated by 
$f_{dc}/f_{th} = 0.95$ (Fig.~5, top), 
$V$ saturates to a finite average velocity after
an initial decay, even though $f_{dc}<f_{th}$. Again in this case the
initial decay is due to individual charge rearrangements and the long
time dynamics corresponds to some channels flowing across the
entire array. 
This saturation occurs due to the finite 
size of the system. Since we have periodic boundary conditions it is possible
for a channel to wrap around the entire system and become stabilized. 
For smaller systems the saturation occurs at lower $f_d$.
in Fig.~6(b) we show the flow channels for the saturation regime 
where a river flows across the system.  

In the experiments \cite{Morgan}, the decaying response persisted
for more than five orders of magnitude in time. In the simulations 
we are limited by both the simulation time and 
by finite size effects.  
In the power law regime at intermediate drives, the flow occurs in 
decreasing numbers of channels
as previously seen in simulations \cite{Reichhardt}.  
In the transient experiments \cite{Morgan} it was speculated that the 
power law decays may arise due to the Coulomb interactions between dots.
We have also considered the case where there is no Coulomb interaction
between mobile charges, and find only exponential or stretched exponential
decays of the conduction.  In addition, the 

\begin{figure}
\center{
\epsfxsize=3.5in
\epsfbox{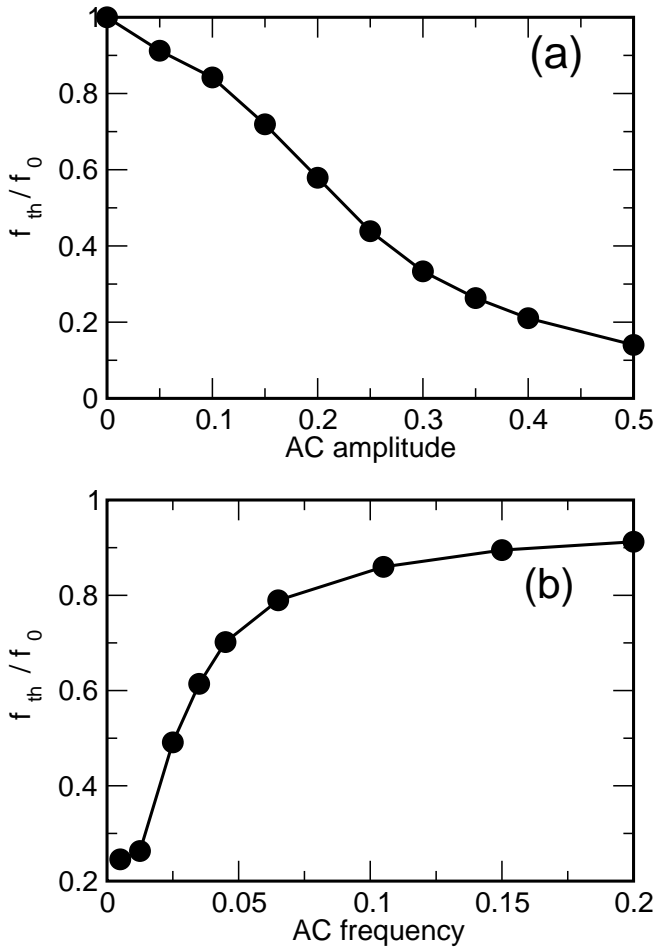}}
\caption{  
(a) The threshold $f_{th}$ for the system in Fig.~1 normalized by 
$f_{0}$, the threshold at zero ac drive, vs ac amplitude for a fixed
ac frequency of $\omega=0.05$. 
(b) The threshold vs ac frequency for a fixed ac amplitude of $A=0.2$.}
\label{fig:ac}
\end{figure}

\noindent
channel structures are not
present,
indicating that the flow through interacting channels plays an important
role in the power law decay of the response.

\section{ac effects} 

We next consider the case $T = 0.0$ for the sample shown in 
Fig.~\ref{fig:scale} 
and measure I-V characteristics for the sample
in the presence of a fixed frequency ac drive.
In Fig.~\ref{fig:ac}(a) we plot $f_{th}/f_{0}$ as a function of ac
amplitude $A$ for fixed frequency $\omega=0.05$,
where $f_{0}$ is the threshold for zero ac drive. 
Here $f_{th}$ monotonically 
decreases for increasing ac amplitude $A$, 
but does not follow a simple functional 
form. 
This decrease is reasonable since, during the positive cycle of the
ac drive, both the ac and dc drives combine to push the charge over
the barrier.  We note
that $f_{0} = 0.14$, indicating that ac amplitudes that are considerably
higher than $f_0$, $A>f_{0}$, still preserve a finite dc threshold
force $f_{th}$. 
We next fix the ac amplitude to $A=0.2$ 
and plot the dependence of $f_{th}$ on the frequency $\omega$
in Fig.~\ref{fig:ac}(b). 
Here, the threshold increases with increasing ac frequency,
with $f_{th}$ saturating as 
$1/\omega$ to $f_{0}$ at the highest frequencies. At
high frequencies, the mobile charge carriers do not have time to respond
to the ac drive. 
We have also measured the scaling of the current-voltage curves, and find that
it is independent of both the ac amplitude and frequency. 
The range of the scaling is, however, reduced by the ac drive.

\section{Summary}

In summary, we have investigated transport in 2D metallic dot
arrays for finite temperature and ac drives.
For zero ac drive and varied temperature, we find a
finite temperature conductance threshold 
which decreases linearly with temperature.
Additionally, the I-V curves obey power law scaling 
with $\zeta = 2.0$, which is independent of the
temperature below a threshold temperature.
These results are in excellent agreement with recent experiments.
For a sudden applied dc drive less than the threshold drive, we find a
two stage decay of the velocity response that shows first
a fast short time 
decay that does not fit to a power law.  This corresponds to 
charge rearrangements less than a lattice constant. 
For longer times
there is a slower long
time decay that is consistent with a power law where the flow consists
of correlated channels that gradually stop. For higher drives that are still
below the threshold, some of the channels can move across the entire
sample and become stabilized. 
If the
long range Coulomb interaction is removed we observe only a fast 
exponential decay.   
We have also studied the effect of superimposing an ac drive on the 
dc drive and find that, for fixed frequency and increasing ac amplitude, 
the threshold decreases. 
Conversely, for fixed amplitude, the 
threshold decreases for decreasing ac frequency. 
The scaling of the current-voltage curves is independent of the ac amplitude
and frequency; however, the range of the scaling changes.

We thank H. Jaeger, R. Parthasarathy, and C. Kurdak 
for useful discussions. 
This work was supported by the US Department of Energy
under Contract No. W-7405-ENG-36.

\end{document}